\documentclass{PoS}
\usepackage[square,sort,comma,numbers]{natbib}
\usepackage{color}
\usepackage{sidecap}

\title{Multiwavelength observations of the blazar BL Lacertae in June 2015}

\ShortTitle{VHE gamma-ray flare of BL Lacertae in June 2015}

\author{\speaker{Shimpei Tsujimoto}\\ 
     Tokai University, Department of Physics,Hiratsuka, Japan\\ 
       E-mail: \email{shimpei.tsujimoto@gmail.com}}

\author{Monica Vazquez Acosta\\ 
     Inst. de Astrofisica de Canarias, E-38200 La Laguna, Tenerife, Spain, Universidad de La
Laguna, Dpto. Astrofisica, E-38206 La Laguna, Tenerife, Spain\\ 
       E-mail: \email{monicava@iac.es}}

\author{Elina Lindfors\\ 
     Tuorla Observatory, Department of Physics and Astronomy, University of Turku, Finland\\ 
       E-mail: \email{ elilin@utu.fi}}

\author{Daniel Mazin\\ 
     Institute for Cosmic Ray Research, the University of Tokyo, Japan\\
     Max Planck Institute for physics, Munich, Germany\\
       E-mail: \email{mazin@mpp.mpg.de}}

\author{Giovanna Pedaletti\\
     Deutsches Elektronen-Synchrotron (DESY) Zeuthen, Zeuthen, Germany\\ 
       E-mail: \email{giovanna.pedaletti@desy.de}}

\author{Vandad Fallah Ramazani\\
     Tuorla Observatory, Department of Physics and Astronomy, University of Turku, Finland\\
       E-mail: \email{vafara@utu.fi}}

\author{Filippo D'Ammando\\
     INAF Istituto di Radioastronomia, Bologna,
Italy\\ 
       E-mail: \email{dammando@ira.inaf.it}}

\author{Julian Sitarek\\
       University of Lodz, PL-90236 Lodz, Poland\\ 
       E-mail: \email{jsitarek@uni.lodz.pl}}

\author{Junko Kushida\\ 
       Tokai University,Department of Physics,Hiratsuka, Japan\\ 
       E-mail: \email{kushida@tokai.ac.jp}}

\author{Kyoshi Nishijima\\ 
     Tokai University,Department of Physics,Hiratsuka, Japan\\ 
       E-mail: \email{kyoshi@tokai-u.jp}}
 
\author{for the MAGIC and Fermi-LAT Collaborations}

\abstract{BL Lacertae is a blazar at the redshift of z = 0.069, eponym of the BL Lac blazar type.
It is also a prototype of the low-frequency-peaked BL Lac (LBL) subclass.
It was first detected in sub-TeV gamma-ray range by MAGIC in 2005. In 2015, MAGIC observations of BL Lacertae were triggered by the {\em Fermi}-LAT analysis report in the MAGIC group, and were performed during 10 individual nights between 15th and 28th June for a total of 8.6 h.
The measured gamma-ray flux varied from 40\% to 10\% of the Crab nebula flux above 200 GeV in the nights from 15th to 17th June.
In particular, a fast variability was found during the nights of 15th and 17th June.
We also performed multi-wavelength (MWL) observations in the radio, optical, UV, X-ray and gamma-ray bands, and 
the MWL light curves indicate that no apparent simultaneous activity in other wavebands accompanying the very high energy gamma-ray flare in June 2015 like an another occurrence of an orphan very high energy flare.
In this proceedings we will present the results of the campaign and discuss their implications on our understanding of the object.
}

\FullConference{35th International Cosmic Ray Conference --- ICRC2017\\
		10--20 July, 2017\\
		Bexco, Busan, Korea}

\begin{document}

\section{Introduction}
Blazars are radio-loud active galactic nuclei (AGN) with the relativistic jets closely aligned to the line of sight of the observer.
This means that it is possible to observe very high energy (VHE: E $>$ 100\,GeV) gamma rays which are emitted from the beamed jets.
Blazars show two bump structures on the spectral energy distributions (SEDs).
The low-energy spectral peak at optical/X-ray range is commonly associated with the synchrotron emissions from relativistic electrons.
On the other hand, the spectral peak at high-energy (HE: E $>$ 100\,MeV) or VHE gamma-ray range is 
usually interpreted as produced by inverse Compton (IC) scattering of the energetic electrons in the jet with synchrotron photons which were emitted by the very same electron population in the synchrotron self-Compton (SSC) scenario,
and/or IC scattering with the thermal photons from outside of the jet in the external Compton (EC) scenario.
Based on their optical spectra, blazars are divided into two classes: flat spectrum radio quasars (FSRQ) that show broad emission lines, and BL Lacertae objects (BL Lacs) characterized by the weakness or even absence of such emission lines (\cite{Weymann1991}, \cite{Stickel1991}).
Depending on the frequency of the low-energy peak of the
SED, the latter class is subdivided into high- (HBL), intermediate- (IBL), and low- (LBL) frequency-peaked BL Lac objects \cite{PadovaniGiommi1995}.

BL Lacertae is a prototype of the BL Lac objects at a redshift of z = 0.069 \cite{MillerJS1978},
and is classified as a LBL.
The first detection of VHE gamma rays from BL Lacertae was reported by the Crimiean Observatory with 7.2$\sigma$ significance above 1\,TeV \cite{Neshpor2001}.
However, this detection is disputed since the object was observed with HEGRA Cherenkov telescopes (which were much more sensitive than the Crimean observations) in the same time and no detection was achieved \cite{Kranich2003}.
After completion of the first MAGIC telescope, BL Lac was observed with the Major Atmospheric Gamma-ray Imaging Cherenkov (MAGIC) telescopes for 22.2\,h in 2005 and for 26 h in 2006, and VHE gamma-rays were discovered in the 2005 data with an integral flux of 3\% of the Crab Nebula flux above 200 GeV \cite{Albert2007}.
This energy threshold is much lower than the one claimed by the Crimean observatory.
The 2006 data of MAGIC observation showed no significant excess.
On 2011 June 28, a very rapid TeV gamma-ray flare 
from BL Lacertae was detected by VERITAS. 
The flaring activity was observed during a 34.6 minute exposure, when the integral flux above 200 GeV reached 
$(3.4\pm0.6)\times 10^{-6}$\,photons m$^{-2}$ s$^{-1}$, roughly 125\% of the Crab Nebula flux \cite{Arlen2013}.

In June 2015, MAGIC observation of BL Lacertae was triggered by the {\em Fermi} Large Area Telescope (LAT) analysis report in the MAGIC group, and the detection of VHE gamma-ray flare
was announced in the Astronomer's Telegrams \#7660.

\section{Observations}
The MAGIC telescopes are a system of two 17\,m Imaging Atmospheric Cherenkov Telescopes (IACTs) located on the Canary island of La Palma, Spain, at $\sim$2200\,m above sea level.
The low energy threshold of the MAGIC telescopes (the standard trigger threshold at low zenith angles is $\sim$50\,GeV) is 
significantly favors observations of VHE gamma rays from extragalactic sources such as blazars because their TeV emission is strongly suppressed by the EBL absorption \cite{Ahnen2015, Ahnen2016}.

The MAGIC observations of BL Lac 
were performed during 10 individual nights between 15th and 28th June 2015 (MJD 57188-57201) for a total of 8.58\,h.
The data were taken with zenith angles in the range 14$^\circ$ to 32$^\circ$.
In addition, the data were taken in the wobble mode \cite{Fomin1994}
meaning that two telescopes alternated four sky positions in every 20 minutes with an offset of 0.4$^\circ$ from the source.

During the 2015 campaign BL Lac was also observed by other instruments at different wave bands.
HE gamma-ray observations were performed by the {\em Fermi}-LAT which is a pair-conversion telescope operating from 20 MeV to $>$ 300 GeV. The {\em Fermi}-LAT data used in this research were collected from 2015 May 1 (MJD 57143) to July 31 (MJD 57234).

X-ray, ultraviolet (UV) and optical observations were performed by the {\em Swift} satellite \citep{Gehrels2004},
which carried out 31 observations of BL Lacertae between 2015 May 2 and July 29.
These observations involved all three instruments on board: the X-ray Telescope [XRT; 0.2--10.0 keV]  \cite{Burrows2005}, 
the UV/Optical Telescope [UVOT; 170--600 nm]\cite{Roming2005}
and the Burst Alert Telescope [BAT; 15--150 keV]\cite{Barthelmy2005}.

Optical observations in R-band were performed as part of the Tuorla blazar monitoring program. 
The observations were performed using a 35\,cm Celestron telescope solidal to the 60\,cm KVA (Kungliga Vetenskapsakademi) Telescope, located at La Palma.
Optical polarization was obtained with Nordic Optical Telescope (NOT) at La Palma and Steward Observatory Telescopes. The NOT observations were performed as part of the dedicated observing program to support MAGIC blazar observations.

BL Lac was quasi-simultaneously observed also by several radio telescopes. Owens Valley Radio Observatory (OVRO) observed the source at 15GHz.
The observations program and the data analysis are described in \cite{Richards2011}.
The Mets\"ahovi radio telescope is a 13.7 meters in diameters that observers at 37 GHz and it is located in Kylm\"al\"a, Finland.
The Boston blazar monitoring program uses Very Long Baseline Array Telescopes to perform monthly monitoring of a sample of blazars at 43 GHz.

\section{Results}
\begin{SCfigure*}
  \centering
  \includegraphics[width=9cm]{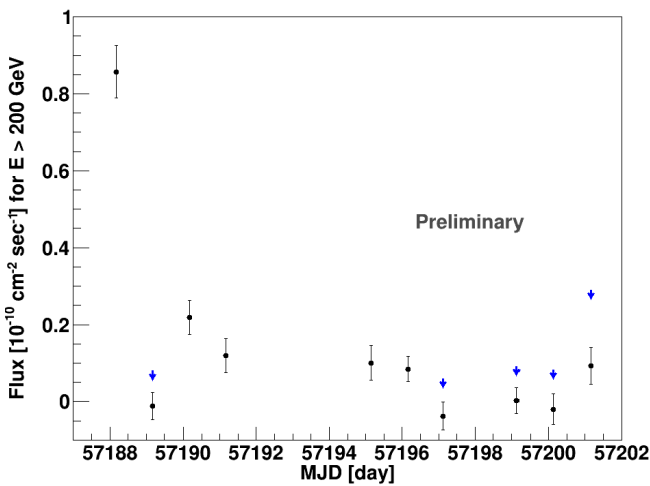}
     \caption{Night-wise light curve of the VHE gamma-ray emission of BL Lacertae above 200\,GeV (black points) including the 95\% upper limits when detection is below 2 sigma (blue arrows). 
     }
  \label{fig:MAGIC_LC}
\end{SCfigure*}
Figure \ref{fig:MAGIC_LC} shows the VHE gamma-ray light curve of BL Lacertae above 200\,GeV during the MAGIC observations.
The analysis results in a signal at 16$\sigma$ significance (according to Li$\&$Ma significance \cite{LiMa1983}).
In particular, the source was detected at 25$\sigma$ significance on 15 June 2015.
The measured VHE gamma-ray flux varied between 40\% and 10\% of the Crab Nebula flux above 200 GeV during the nights of 15th and 17th June, respectively.

\begin{SCfigure*}
  \centering
  \includegraphics[width=11cm]{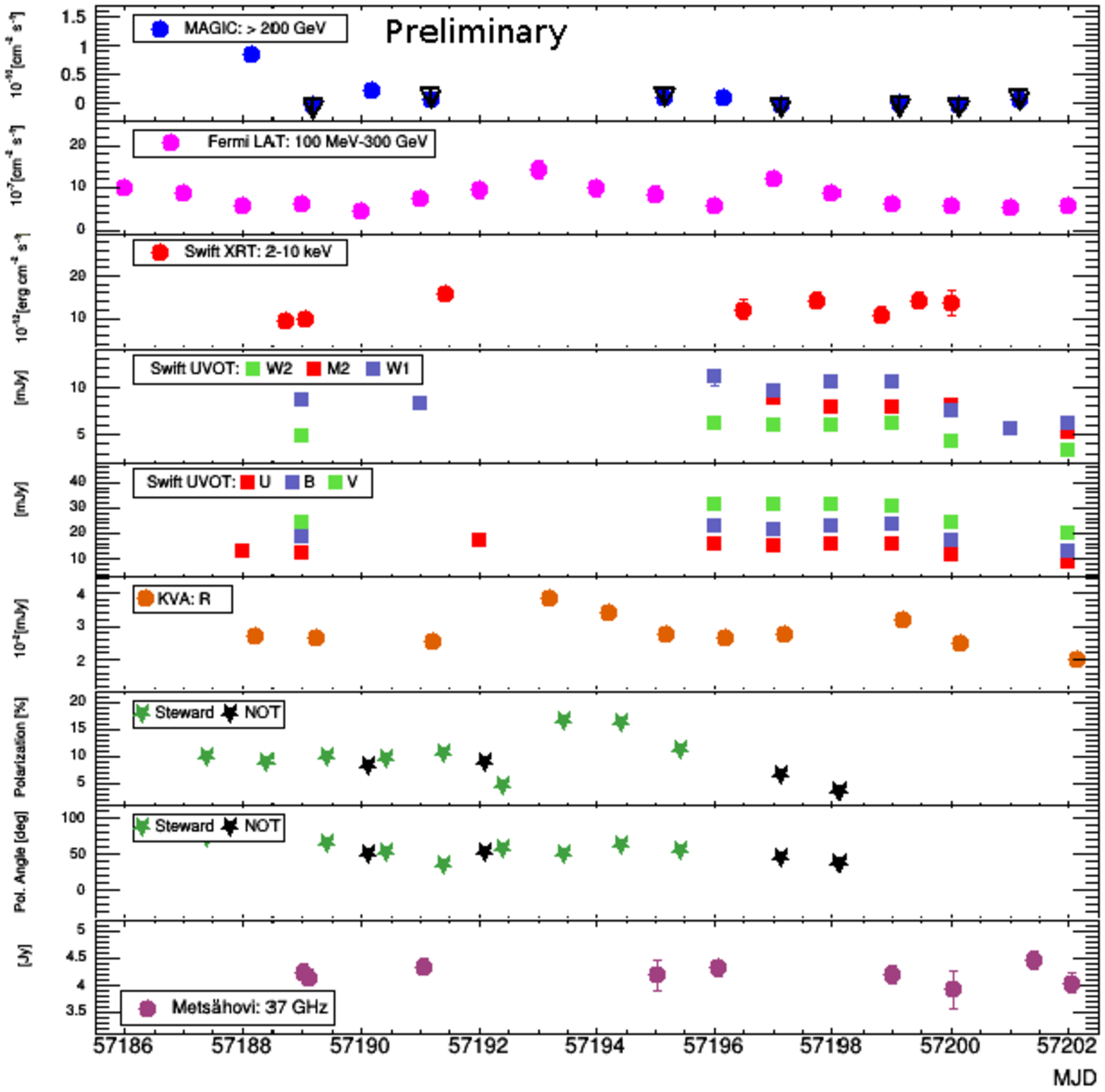}
     \caption{Multiwavelength light curve of BL Lacertae (from top to bottom: VHE gamma rays, MAGIC; HE gamma rays, {\em Fermi}-LAT; X rays, {\rm Swift}-XRT; UV and optical, {\rm Swift}-UVOT; optical, KVA; optical polarization, Steward and NOT; Radio, Mets\"ahovi and OVRO) in June 2015.}
  \label{fig:MWL_LC}
\end{SCfigure*}

Figure \ref{fig:MWL_LC} shows the multiwavelength light curve of BL Lacertae during the MAGIC observation period (MJD 57188-57201).
The HE daily light curve (by {\em Ferm}-LAT, second from the top) shows at least one prominent peak around MJD 57193.
The X-ray flux is varied without a clear outburst.
UV W2, M2, W1 bands are smoothly variable during the period.
The optical polarization degree and angle were also changing, especially around the optical high state observed in R-band,
but no strong variation and rotation are observed.
There is an apparent correlation between variability in optical and UV bands and the one in the HE gamma rays.
The radio light curves show no strong outburst during the period.
However, there is no strong flare in radio to HE gamma rays at the time of the gamma-ray flare (MJD 57188).
Therefore, this VHE gamma-ray flare seems to be a so-called orphan flare.
This behaviour is similar to the one observed in the fast flare in 2011 \cite{Arlen2013}.

\begin{SCfigure*}
  \centering
  \hspace{1.5cm}
  \includegraphics[width=9cm]{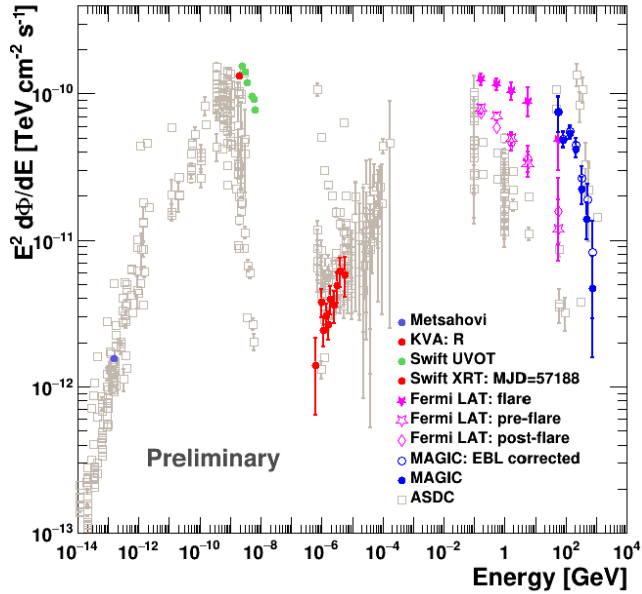}
  \caption{Multiwavelength SED of BL Lacertae.  
   Data for each instrument are collected in the following period: Mets\"ahovi, KVA, {\em Swift}-UVOT and {\em Fermi}-LAT flare: MJD 57188-57201.
   {\em Swift}-XRT and MAGIC: MJD 57188. 
   {\em Fermi}-LAT pre-flare: MJD 57143-57187. 
   {\em Fermi}-LAT post-flare: MJD 57202-57234.
   Grey open circles: archival data from ASDC (http://www.asdc.asi.it/).
   }
  \label{fig:MWL_SED}
\end{SCfigure*}

Figure \ref{fig:MWL_SED} shows the multiwavelength SED.
The MAGIC spectrum is corrected for absorption via pair-production using the model for extragalactic background light (EBL) in \cite{Dominguez2011}.
Hard X-ray and gamma-ray spectra show the broad IC component like FSRQ.
However, the SED is not Compton dominated.
No simultaneous activity at other wavelength is observed from the light curves, and the broad IC component without large Compton dominance might indicate that an additional component could be interpreted in an intermediate state between BL Lac and FSRQ
.

\section{Summary}
In June 2015, a gamma-ray flare was detected with the MAGIC telescopes.
Multiwavelength observations were performed from radio to HE gamma rays during the VHE gamma-ray observations.
There is no apparent simultaneous activity at other wavelengths accompanying the VHE gamma-ray flare in June 2015, which indicates an another occurrence
of an orphan VHE flare.

\section{Acknowledgement}
We would like to thank the Instituto de Astrofisica de Canarias for the excellent working conditions at the Observatorio del Roque de los Muchachos in La Palma. 
The financial support of the German BMBF and MPG, 
the Italian INFN and INAF, 
the Swiss National Fund SNF, the ERDF under the Spanish MINECO (FPA2015-69818-P, FPA2012-36668, FPA2015-68378-P, FPA2015-69210-C6-2-R, FPA2015-69210-C6-4-R, FPA2015-69210-C6-6-R, AYA2015-71042-P, AYA2016-76012-C3-1-P, ESP2015-71662-C2-2-P, CSD2009-00064), 
and the Japanese JSPS and MEXT is gratefully acknowledged. 
This work was also supported by the Spanish Centro de Excelencia "Severo Ochoa" SEV-2012-0234 and SEV-2015-0548, 
and Unidad de Excelencia "Maria de Maeztu" MDM-2014-0369, 
by the Croatian Science Foundation (HrZZ) Project 09/176 and the University of Rijeka Project 13.12.1.3.02, 
by the DFG Collaborative Research Centers SFB823/C4 and SFB876/C3, 
and by the Polish MNiSzW grant 2016/22/M/ST9/00382.

The \textit{Fermi}-LAT Collaboration
acknowledges support for LAT development, operation and data analysis from
NASA and DOE (United States), CEA/Irfu and IN2P3/CNRS (France), ASI and
INFN (Italy), MEXT, KEK, and JAXA (Japan), and the K.A.~Wallenberg
Foundation, the Swedish Research Council and the National Space Board
(Sweden). Science analysis support in the operations phase from INAF
(Italy) and CNES (France) is also gratefully acknowledged. This work
performed in part under DOE Contract DE-AC02-76SF00515.


\end{document}